\documentstyle[12pt,a4]{article}
\author{E.~ L.~ Afraimovich,  E.~ A.~ Kosogorov, O.~ S.~ Lesyuta  \\
        Institute of Solar-Terrestrial Physics SD RAS,\\
        p.~o.~box~4026, Irkutsk, 664033, Russia\\
        fax: +7 3952 462557; e-mail:~afra@iszf.irk.ru}

\title{EFFECTS OF THE AUGUST 11, 1999 TOTAL SOLAR ECLIPSE AS DEDUCED FROM
       TOTAL ELECTRON CONTENT MEASUREMENTS AT THE GPS NETWORK}
\date{}
\begin{document}
\sloppy
\maketitle
\begin{abstract}
We present the results derived from measuring fundamental
parameters of the ionospheric response to the August 11, 1999
total solar eclipse. Our study is based on using the data from
about 70 GPS stations located in the neighbourhood of the eclipse
totality phase in Europe. The eclipse period was characterized by
a low level of geomagnetic disturbance ($Dst$-variation from -10
to -20 nT), which alleviated significantly the problem of
detecting the ionospheric response to the eclipse. Our analysis
revealed a well-defined effect of a decrease (depression) of the
total electron content (TEC) for all GPS stations. The delay
between minimum TEC values with respect to the totality phase
near the eclipse path increased gradually from 4 min in Greenwich
longitude (10{:}40 UT, LT) to 8 min at the longitude $16^\circ$
(12{:}09 LT). The depth and duration of the TEC depression were
found to be 0{.}2-0{.}3 TECU and 60 min, respectively. The
results obtained in this study are in good agreement with earlier
measurements and theoretical estimates.
\end{abstract}

\section{Introduction}
\label{TSE-sect-1}

Experimental observations of the ionosphere at the time of solar eclipses
provide the source of information about the character of behavior of  the
various  ionospheric  parameters.  Regular  ionospheric  effects of solar
eclipses are fairly well understood.  They imply an increase of effective
reflection heights,  a reduction in concentration in the F-layer maximum,
and a decrease in total electron content (TEC) in the  ionosphere,  which
is  typical of the transition to the nightside ionosphere (Cohen, 1984).
The behavior of  the  above  parameters  can  be  modeled  using
appropriate ionospheric models (Stubbe, 1970; Boitman et al., 2000).

The main parameters of the ionospheric response  are  the  value  of  the
delay  $\tau$ with respect to the eclipse totality phase,  as well as its
amplitude $A$ and duration $\Delta T$. Almost all publications devoted to
the study of the ionospheric response to solar eclipses present estimates
of these parameters.  A knowledge of these values makes  it  possible  to
refine, in terms of the respective aeronomic ionospheric models, the time
constants of ionization and recombination processes at different  heights
in the ionopshere.

The measurements  of $\tau$ were made by analyzing the characteristics of
the ionosphere-reflected radio signal at vertical-incidence soundings  at
a  network  of  ionospheric  stations  (Marriott  et  al{.},  1969;
Goncharov  et  al{.},  1982; Boitman  et  al{.},  2000)   by
measuring  the frequency Doppler shift at vertical- and oblique-incidence
soundings (Boitman et al{.},  2000). In the cited references, the
value of $\tau$ was found to vary from 0 (Marriott et al., 1969)
to 20 min, with the amplitude $A$ of a decrease in local electron density
of  order  $9  \times  10^{4}$ cm${}^{-3}$,  and the response duration of
about 1 hour (Boitman et al., 2000).

Similar measurements at an ionospheric station were made during the solar
eclipse of September 23, 1987 in the south-eastern Asia (Cheng, 1992).
The amplitude,  delay and duration of the ionospheric response were found
to  be  $0{.}3  \times  10^{6}$  cm${}^{-3}$,  18  min and 1 hour 30 min,
respectively.  At the same  time,  measurements  during  the  eclipse  of
October  24,  1995  in  the same region provided an estimate of $\tau$ of
about 80 min (Huang,  1999), $\Delta T$ = 1 hour 20 min, and $A$ in
excess  of  $1 \times 10^{6}$ cm${}^{-3}$.  The measurements of $f_{0}F2$
over the Scaramanga station during a partial solar  eclipse  of  May  20,
1996  (Anastassiadis,  1970) give the value of $\tau$ of about 38 min.
In paper Zherebtsov et al. (1998a),  the delay time of  the  ionospheric
response to the eclipse was between 25 and 30 min (at 300 km altitude).

Interesting results  were  obtained  in  observations of an annular solar
eclipse of May 30,  1984 at the Millstone Hill incoherent  scatter  radar
(Salah,  1986).  A decrease in electron density with respect to the
eclipse totality phase occurred in this case 20--30 min later,  with $A$ =
$4 \times 10^{6}$cm${}^{-3}$,  and $\Delta T$ = 2 hours 15 min.  However,
facilities of this kind are too few to be extensively used in experiments
during solar eclipses.

The ionospheric  response  to  the March 9,  1997 total solar eclipse was
investigated using a network of ionosondes  (Zherebtsov  et  al{.},
1998b). For this event, the position of the minimum of the trough
in the time dependence of electron density was shifted with respect to  the
time  of  maximum  occultation  of the solar disk by 10--15 min,  with
$A$ = $1 \times 10^{5}$ cm${}^{-3}$, $\Delta T$ = 1 hour 15 min.

A large amount of data was obtained by measuring the Faraday rotation  of
the  plane  of  polarization  of  VHF signals of geostationary satellites
(Klobuchar et al{.}, 1970; Hunter et al{.}, 1974; Davies,1980;
Rama Rao, 1982; Essex, 1982; Deshpande, 1982; Singh, 1989).  These
measurements revealed the eclipse-induced effect of  a  deep  depression
(decrease) in TEC with the amplitude $A$ varying  from 2 to 14~TECU, and
with a typical time of TEC depression and recovery of about several hours
(the unit of measurement of TEC that is adopted in the literature
corresponds to $10^{16}$ ¬${}^{-2}$).  The  spread  in  the values  of
$\tau$ was also found to vary over a wide range, from 5 to 40 min.

According to Klobuchar et al{.} (1970), for a total solar eclipse
that occurred in spring (March 7, 1970), the amplitude $A$ was
11~TECU, $\Delta T$ = 2 hours 44 min, and $\tau$ of order 33 min.
For the total solar eclipse of February 16, 1980, the
observations at Waltair (Singh, 1989) gave the following
estimates: $A$ = 0{.}11~TECU, $\Delta T$ = 1 hour, and $\tau$ =
20 min. Essex et al. (1982) inferred that for the total solar
eclipse of February 20, 1979, the values of $A$ and $\tau$ are
14~TECU and 33~min, respectively.

Hence a  large  body  of  experimental  data do not permit us to make any
reliable estimates of the basic parameters of the  ionospheric  response.
One  of the reason for such a great difference is attributable to the use
of different methods of measurements which differ greatly by spatial  and
temporal  resolution.  However,  the  main reason is caused by dissimilar
characteristics of the  eclipse  itself,  by  geophysical  conditions  of
individual  measurements,  and  by  a  large  difference of the latitude,
longitude and local time when experiments are conducted.

To obtain more reliable information about the behaviour of the ionosphere
during an eclipse, it is necessary to carry out simultaneous measurements
over  a  large  area  covering  regions  with  a  different  local  time.
Furthermore,  high  spatial  (of  some  tens  of kilometres at least) and
temporal (at least 1 min) resolution is  needed.  However,  none  of  the
above familiar methods meets such requirements.

The development of the global navigation system GPS and the creation,  on
its basis,  of extensive networks of GPS stations (which at  the  end  of
1999 consisted of no less than 600 sites), the data from which are placed
on the INTERNET (Klobuchar, 1997),  open up a new era  in  remote
sensing  of  the ionosphere.  At almost any point of the globe and at any
time  at  two  coherently-coupled  frequencies  $f_1=1575{.}42$  MHz  and
$f_2=1227{.}60$  MHz,  two-frequency  multichannel  receivers  of the GPS
system are used to carry out high-precision measurements of the group and
phase  delay along the line o sight between the ground-based receiver and
satellite-borne transmitters in the zone of  reception.  The  sensitivity
afforded  by  phase measurements in the GPS system permits irregularities
to be detected with an amplitude of up  to  $10^{-3}$--$10^{-4}$  of  the
diurnal variation of TEC.

Afraimovich et al. (1998) were the first to use
in a detailed analysis of regular ionopsheric effects measurements of TEC
and its gradients for the total solar eclipse of March 9, 1997, using for
this purpose the  GPS  interferometer  at  Irkutsk.  Their  results  bear
witness to significant changes occurring in the process of ion production
in the ionosphere during the solar eclipse,  simultaneously  in  a  large
volume of space with a radius of at least 300 km at 300 km altitude.  The
delay of a minimum TEC value with respect to the  phase  of  totality  is
about  10 min,  the depth of TEC depression varies from 1 to 3~TECU,  and
$\Delta T$ = 1 hour 15 min.

Boitman et al. (2000) obtained a good agreement
of  these  data,  as well of measurements for several HF Doppler sounding
paths   with   results   of   a   numerical   simulations  using    the
ionosphere-plasmasphere  coupling model (Krinberg, 1984). Unfortunately,
during this eclipse the total shadow's path in 1997 was monitored  by  only
few  GPS stations; therefore, it was not possible  to   determine   the
spatial-temporal characteristics of the ionopsheric response.

According to the data from five GPS stations  reported  by  Tsai  and
Liu (1999) for the solar eclipses of October
24, 1995 and March 9, 1997,  the delay $\tau$ varied over a broad  range,
from 0 to 30 min,  the amplitude $A$ was close to 7~TECU,  and $\Delta T$
varied from 40 min to 1 hour.

A unique opportunity to exploit the potential  of  the  GPS  network  was
provided by the total solar eclipse of August 11,  1999.  For this period
of time,  the INTERNET made available the data  from  at  least  100  GPS
stations  located  in  Western  and  Central  Europe  within and near the
totality path.  A massive flow of publications on the various ionospheric
effects  of this eclipse should be forthcoming,  based on the GPS data in
conjunction with the data from other observing facilities.  The objective
of  this  paper  is  to determine the basic parameters of the ionospheric
response to the August 11, 1999 total solar eclipse using these data.

General information about the eclipse, and a description of the
experimental geometry are given in Section~\ref{TSE-sect-2}. The
ionospheric response to the eclipse is discussed in
Section 3 using the data from reference
ionospheric station Chilton. The processing technique for the GPS
network data, and results derived from analysing the ionospheric
response of the August 11, 1999 solar eclipse in Europe are
outlined in Section 4. Results obtained in this
study are discussed in Section 5.

\section{The geometry and general  information  of  total  solar  eclipse
AUGUST 11, 1999}
\label{TSE-sect-2}

The last solar eclipse in the 20--th century began in North Atlantic, and
path of the Moon's shadow made first landfall in south-western England at
10{:}10~UT;  the  Sun at that time was at an angle of $45^\circ$ over the
eastern horizon.  The centre line duration of total  eclipse  averaged  2
min, and the total eclipse was confined to a narrow corridor 103 km wide.

Fig.~1 shows a schematic map of the path of the Moon's shadow
crossing parts of Western and Central Europe (the data from  Espenak,
1999  were used in constructing this map).  The centre line of eclipse at
ground level is shown as a thick line,  and thin lines correspond to  its
southern  and northern boundaries.  The location of reference ionospheric
station Chilton (RAL) is marked by the symbol $\star$. Dots $\bullet$ and
symbols  $+$ show the locations of the GPS stations used in the analysis;
their geographic coordinates are presented in Table~1.
Dots $\bullet$, together with symbols $+$ are used to represent a total
set of GPS stations. Symbols $+$ form a network of GPS stations located
near the eclipse path; therefore, we designated this  group as the near
zone. Numbers for the longitudes $10^\circ  W$,  $0^\circ$,  $10^\circ E$,
$20^\circ E$,  $30^\circ E$, $40^\circ E$ correspond to the local time
for these longitudes.

In this paper we confine ourselves to  analysing  only  to  a  region  of
Western  and  Central  Europe  from  the coast of southern England to the
point with coordinates $56{.}03^\circ N$,  $37{.}2^\circ  E$,  where  the
totality  phase  was observed at 11{:}20~UT (14{:}20~LT).  Thus the solar
eclipse  effect  occurred  for  the  conditions  of  the  daytime  summer
ionosphere.

The distance  along  the  great-circle  arc  between  the above-mentioned
extreme points is about 2900 km, with the time difference of 67 min only.
Hence  a distinguishing characteristic of this eclipse was the supersonic
speed of the Moon's shadow  sweeping  through  the  terrestrial  surface,
exceeding 720 m/s.

At ionospheric  heights the totality path was travelling somewhat further
south.  The onset time of the totality phase for the height h=300 km over
Budapest  is  1{.}3 min ahead of that at ground level.  The difference in
the values of the totality phases and their onset times is caused by  the
Sun's  altitude  over  the horizon.  At the time of the totality phase in
Budapest (11{:}05~UT, or 14{:}05~LT), it was as small as $59^\circ$.

The period under consideration was characterized by a low level
of geomagnetic disturbance ($Dst$-variation from -10 to -20 nT),
which simplified greatly the problem of detecting the ionospheric
response of eclipse.

\section{The ionospheric response by eclipse  from  date  of  ionospheric
station Chilton}
\label{TSE-sect-3}

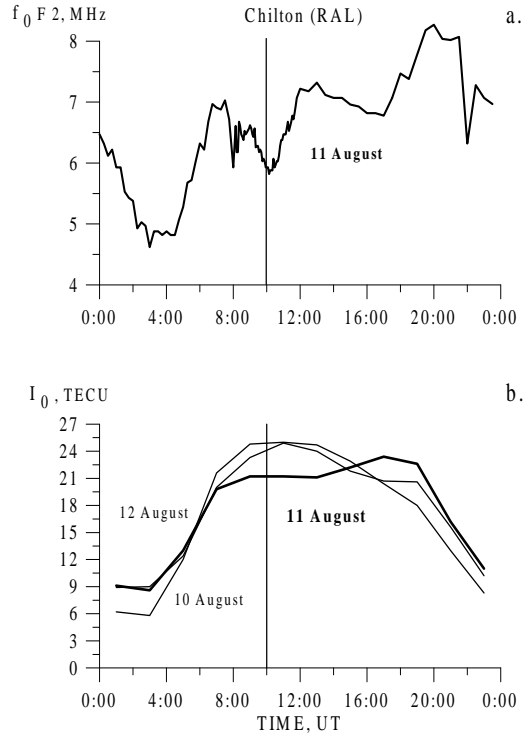
First we consider the variations of critical frequencies
$f_{0}F2$ over the time interval 00{:}00-24{:}00~UT on August 11, 1999
according to the data from station Chilton (RAL) - Fig.~2a.
The onset time of the totality phase of eclipse in the area of
station Chilton (RAL) at 300 km altitude is shown by a vertical solid
line. As might be expected under the conditions of the summer
ionosphere, the mean level of $f_{0}F2$ differs only slightly for the
night-time and daytime. Nevertheless, a decrease of $f_{0}F2$ during
the totality phase of eclipse is sufficiently clearly
distinguished.

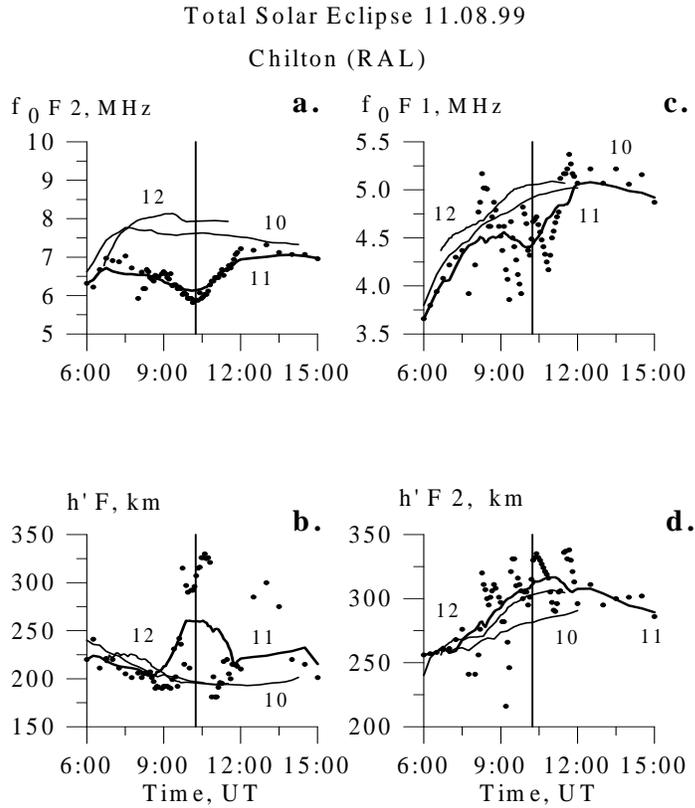
Consider  the variations of ionospheric parameters for the time
interval 06{:}00--15{:}00~UT on August 11,  1999,  and on the  background
days  of  August  10 and 12,  using the data from station Chilton (RAL) -
Fig.~3. Dots correspond to variations of critical frequencies
$f_{0}F2$ (panel  a),  $f_{0}F1$ (panel c), apparent heights $h'F$
(panel b), and $h'F2$ (panel d) for August 11,  1999.  Solid curves plot
the same values that  are  smoothed  with a time window of 60 min.  For
August 10 and 12, only smoothed curves are given,  with the same time window
of 60 min. The onset  time  of  the  totality  phase  of  eclipse
(10{:}16~UT) at 300 km altitude over the station is shown by a thin
vertical line.

The eclipse effect is most conspicuous  in  the  variations  of  critical
frequencies  $f_{0}F2$,  whose  maximum  difference  from  the background
values on August 10 and 12 at the time of reaching a minimum (10{:}20~UT)
was  up  to  2  MHz.  On  the other hand,  the amplitude of a decrease of
$f_{0}F2$ for the August 11 event (after the totality phase  of  eclipse)
does  not exceed 1 MHz.  The eclipse effect on other parameters is not as
clearly distinguished,  and  becomes  evident  only  when  comparing  the
smoothed  curves.  Similar  results  of  measurements  at  an ionospheric
station were obtained during the total solar  eclipse  of  September  23,
1987 in South-Eastern Asia (K.~Cheng, 1992).

The $f_{0}F2$  --  variations,  measured at station Chilton (RAL) at time
intervals of   4   min,   are   presented   in   greater   detail   in
Fig.~4a (heavy dots).  The solid curve connecting these dots
is an approximating one for these values.  This panel plots also the
geometrical function of  eclipse $S(t)$ at 300 km altitude, calculated for
station Chilton. Minimum values of $f_{0}F2$ and $S(t)$ correspond to the
points A and B in this figure. The respective delay $\tau$ between the time
of a minimum of $f_{0}F2$ and of the function $S(t)$ is close to 4 min in
this case.

The data from the station Chilton (RAL) and from GPS station HERS nearest
to it are compared in the next Section.

\section{The process  of  GPS--network  data  and  results of analysis of
ionospheric  effect  by  total  solar  eclipse  of August   11,   1999}
\label{TSE-sect-4}

We now  give  a  brief  account  of  the  sequence  of procedures used in
processing the  GPS  data.  Input  data  are  represented  by  series  of
"oblique" values of TEC $I(t)$, as well as by the corresponding series of
elevations $\theta(t)$ measured from the ground, and azimuths $\alpha(t)$
of  the  line  of  sight  to  the  satellite  measured clockwise from the
northward direction.  These parameters are calculated  by  our  developed
CONVTEC  program  by  converting  the  GPS-standard  RINEX-files from the
INTERNET.  Series of elevations $\theta(t)$ and azimuths  $\alpha(t)$  of
the  line of sight to the satellite are used to determine the location of
subionospheric points.  In the case under consideration, all results were
obtained for larger than $45^\circ$ elevations $\theta(t)$.

Fig.~4c presents the experimental geometry in the area of
ionospheric station Chilton (RAL) - $\star$, and GPS station HERS
($\bullet$). Heavy dots $\bullet$ show the centre line of eclipse at ground
level, and smaller dots correspond to its southern and northern boundaries.
The symbol $\ast$ shows the position of the subionospheric point at the time
of a maximum response of TEC (see below).

Various methods  for  reconstructing  the  absolute  value  of  TEC   are
currently  under  development  using measurements of both phase and group
delays;  however,  effective algorithms for an accurate solution of  this
problem  are  still  unavailable for the different types of two-frequency
receivers and operating modes of the GPS system.  In this connection, for
purposes of this paper,  we limit our attention to considering only those
TEC variations which were  obtained  from  phase  delay  measurements  by
formula (Afraimovich et al., 1998):

\begin{equation}
\label{TSE-eq-01}
I_p=\frac{1}{40{.}308}\frac{f^2_1f^2_2}{f^2_1-f^2_2}
                           [(L_1\lambda_1-L_2\lambda_2)+const+nL]
\end{equation}

where $L_1\lambda_1$ and $L_2\lambda_2$ are additional paths of the radio
signal caused by the phase delay in the ionosphere,~(m);  $L_1$ and $L_2$
represent the number of phase rotations  at  the  frequencies  $f_1$  and
$f_2$;   $\lambda_1$   and   $\lambda_2$   stand  for  the  corresponding
wavelengths,~(m);  $const$ is the unknown initial  phase  ambiguity,~(m);
and $nL$~ are errors in determining the phase path,~(m).

For an approximate specification of the TEC constant component
$I_0$ at time intervals of 2 hours, we made use of the INTERNET
data on corresponding global maps of the absolute vertical value
of TEC in the IONEX format (Mannucci, 1998) - see also Fig.~2b.
To normalize the response amplitude we converted the "oblique"
TEC to an equivalent "vertical" value:

\begin{equation}
\label{TSE-eq-02}
I(t)=I_p(t)sin(\theta(t))
\end{equation}

Although, because of a strong horizontal TEC gradient and without  regard
for  the  spericity  of  the  problem,  this procedure gives a very rough
result,  but this result is quite acceptable because all our results were
obtained for larger than $45^\circ$ elevations $\theta(t)$.

With the purpose of eliminating variations of the regular ionosphere,  as
well as trends introduced  by  the  satellite's  motion,  we  employ  the
procedure of eliminating the trend by preliminarily smoothing the initial
series with the time window in the range from 40  to  100  min  which  is
fitted  for  each  TEC sampling.  Such a procedure is also required for a
clearer identification of the ionospheric response of  eclipse  which  is
characterized  by  a  relatively  small  amplitude  (see below) under the
presence of space-time TEC variations that are not  associated  with  the
eclipse.

For the purposes of illustration of the data processing procedure
Fig.~5 presents the filtered TEC variations $dI(t)$
for August 11 (thick line), and for the background days of August
10 and 12, 1999 (thin lines) for station HERS for satellite N14
(PRN 14) - c), as well as for stations ZIMM (PRN 1) - b) and WTZT
(PRN 14) - a), which are separated from station HERS,
respectively, by $7^\circ$ and $13^\circ$ in longitude eastward.
Figures correspond to GPS satellite numbers. The onset time of
the totality phase of eclipse at 300 km altitude for the
corresponding subionospheric points is shown by vertical solid
lines. As is evident from the figure, responses to the eclipse at
these stations are very similar in both form and amplitude, but
the response delay increases with the longitude. $dI(t)$ --
variations for the background days of August 10 and 12, 1999 for
different stations differ substantially not only from the TEC
response to eclipse but also from one another.

Fig.~4b presents the filtered variations of TEC $dI(t)$ for
station HERS for PRN 14 during August 11, 1999 (thick line). This panel
plots also the geometrical function of eclipse at 300 km altitude, $S(t)$,
that is calculated for the subionospheric point of PRN 14. A minimum value
of $S(t)$ corresponds to the point A in the figure.

As is apparent from this figure, the form of the filtered
variations is similar to a triangle whose vertex (point B)
corresponds to the time at which the TEC attains its minimum
value. The value of $dI_{min}$ itself can serve as an estimate of the
amplitude of TEC response to eclipse, and the time interval
between the times of intersection of the line $dI$=0 (points C
and D) can serve as an estimate of the duration of the response
$\Delta T$. The corresponding delay $\tau$ between the times of the
minimum of $dI_{min}$ and of the function $S(t)$ in this case was found
to be 4 min, which coincides with the estimate of $\tau$ for
ionospheric station Chilton (see Section ~\ref{TSE-sect-3}). The response
amplitude in this case was close to 0{.}18 TECU, and the response
duration was 46 min.

Such $dI(t)$ -- variations are characteristic for all GPS
stations and satellite numbers 01 and 14 listed in Table~1.
The choice of the same satellites, PRN 01 and PRN 14, for the entire
selected set of GPS stations was dictated by the fact for these
satellites a maximum value of the elevation $\theta$ of the line of
sight to the satellite exceeded $45^\circ$ for the time interval
10{:}00-12{:}30~UT, which reduced to a minimum the possible error of
conversion to the "vertical" value of TEC as a consequence of
the sphericity.

The first line of Table~2  presents the results of a
statistical processing for the entire set of GPS stations listed
in Table~1. The second line includes only those stations
whichlie in the immediate vicinity of the eclipse path, within $\pm$
$5^\circ$ with respect to the centre line (near zone). In Table~2,
the values before the bar are mean values, and those after the bar
correspond to the standard deviation. The mean value of $\tau$ for
the entire set of stations is 16 min, while for the stations in
the near zone this value is 7 min. The mean value of the
amplitude A=0{.}3 TECU for all stations and A=0{.}1 TECU for the
near zone. The width of the TEC trough for the far and near
zones $\Delta T$ = 60 min, on average.

Fig.~6 presents a longitudinal dependence of the time
position $t_{min}$ of minima of the curves $dI(t)$ for subionospheric
points lying in the immediate vicinity of the eclipse band
(satellites 1 and 14) - heavy dots. Dark symbols $\triangle$
designate the times of totality phases of the geometrical
function of eclipse versus longitude which are calculated for
the subionospheric points. Shaded symbols $\diamond$ correspond to
delay times between maximum of the geometrical function of
eclipse and a minimum in TEC.

It was found that the delay $\tau$ increases gradually from 4
min at the Greenwich longitude (10{:}23~UT, LT) to 16 min at the
longitude of $16^\circ$ (12{:}09~LT).

\section{Conclusions}
\label{TSE-sect-5}

Our results are in good agreement with earlier measurements
and theoretical estimates (see a review of the data in the
Introduction). The key feature of our data is a higher
reliability of determining the main parameters of the response
to eclipse which is due to high space-time resolution and to the
increased sensitivity of detection of ionospheric disturbances
inherent in the GPS-array method which we are using.

Also, due regard must be had to the fact that the distinctive
property of the eclipse under consideration was a relatively
small response amplitude, which required special filtering of the
TEC series (see preceding Section). The reason is that, unlike a
number of eclipses for which more-or-less reliable data were
obtained, this eclipse occurred in the summer season
characterized by only moderate differences of the daytime and
night-time ionization. Furthermore, in this situation the time
variation of $f_{0}F2$ and vertical TEC near noon usually shows a
minimum which, in essence, masks the eclipse effect (see also
Fig.~2b). It is also vital to note that the time constant of a
decrease in ionization in the $F_2$ maximum exceeds substantially
the duration of the totality phase of eclipse, which leads also
to a decrease in response amplitude.

The local time-dependence of $\tau$ that is revealed in this
paper is in agreement with theoretical estimates reported in
(Stubbe, 1970). The value of $\tau$ for foF2, approaching 6 min,
corresponded to 13{:}40 LT. Using modeling methods (Ivelskaya et al.,
1977) showed that the variations of the delay time $\tau$ of minimum
local electron density $N_e(t)$ with respect to a minimum of the
ion production function are as follows: $\tau$ = 1-2 min at 150 km
altitude, $\tau$ = 3 min at 200 km, $\tau$ = 20 min at 300 km, and
$\tau$ = 45 min above 600 km. In this paper $\tau$ is estimated,
respectively, at about 3 min for 200 km altitude and 20 min for 300 km
altitude for 12~LT.

\section{Acknowledgements}
\label{TSE-sect-6}

We are grateful to ~K.~S.~Palamartchouk, ~A.~V.~Tashchilin and
A.~D.~Kalikhman for their interest in this study, helpful advice
and active participation in discussions. Thanks are also due to
V.~G.~Mikhalkosky for his assistance in preparing the English
version of the manuscript. This work was done with support from
the Russian foundation for Basic Research (grant 99-05-64753)
and RFBR grant of leading scientific schools of the Russian
Federation No. 00-15-98509.

\end{document}